\documentclass[aps,prb,twocolumn,superscriptaddress,floatfix,longbibliography]{revtex4-2}

\usepackage{amsmath,amssymb} % math symbols
\usepackage{bm} % bold math font
\usepackage{graphicx} % for figures
\usepackage{comment} % allows block comments
\usepackage{textcomp} % This package is just to give the text quote '

\usepackage{blindtext}

\usepackage{enumitem}
\setlist{noitemsep,leftmargin=*,topsep=0pt,parsep=0pt}

\usepackage{xcolor} % \textcolor{red}{text} will be red for notes
\definecolor{lightgray}{gray}{0.6}
\definecolor{medgray}{gray}{0.4}

\usepackage{hyperref}
\hypersetup{
colorlinks=true,
urlcolor= blue,
citecolor=blue,
linkcolor= blue,
}

\newif\ifptitle
\newif\ifpnumber
\newcounter{para}
\newcommand\ptitle[1]{\par\refstepcounter{para}
{\ifpnumber{\noindent\textcolor{lightgray}{\textbf{\thepara}}\indent}\fi}
{\ifptitle{\textbf{[{#1}]}}\fi}}
\newcommand{\mytitle}{Atomic photoexcitation as a tool for probing purity of twisted light modes}

\begin{document}

\title{\mytitle}

\author{R.~P.~Schmidt}
\email[]{riaan.schmidt@ptb.de}
\affiliation{Physikalisch-Technische Bundesanstalt, Bundesallee 100, D-38116 Braunschweig, Germany}
\affiliation{Institut für Mathematische Physik, Technische Universität Braunschweig, Mendelssohnstrasse 3, D-38106 Braunschweig, Germany}

\author{S.~Ramakrishna}
\affiliation{Helmholtz-Institut Jena, Fröbelstieg 3, D-07763 Jena, Germany}
\affiliation{GSI Helmholtzzentrum für Schwerionenforschung GmbH, Planckstrasse 1, D-64291 Darmstadt, Germany}
\affiliation{Theoretisch-Physikalisches Institut, Friedrich-Schiller-Universität Jena, D-07763 Jena, Germany}

\author{A.~A.~Peshkov}
\affiliation{Physikalisch-Technische Bundesanstalt, Bundesallee 100, D-38116 Braunschweig, Germany}
\affiliation{Institut für Mathematische Physik, Technische Universität Braunschweig, Mendelssohnstrasse 3, D-38106 Braunschweig, Germany}

\author{N.~Huntemann}
\affiliation{Physikalisch-Technische Bundesanstalt, Bundesallee 100, D-38116 Braunschweig, Germany}

\author{E.~Peik}
\affiliation{Physikalisch-Technische Bundesanstalt, Bundesallee 100, D-38116 Braunschweig, Germany}

\author{S.~Fritzsche}
\affiliation{Helmholtz-Institut Jena, Fröbelstieg 3, D-07763 Jena, Germany}
\affiliation{GSI Helmholtzzentrum für Schwerionenforschung GmbH, Planckstrasse 1, D-64291 Darmstadt, Germany}
\affiliation{Theoretisch-Physikalisches Institut, Friedrich-Schiller-Universität Jena, D-07763 Jena, Germany}

\author{A.~Surzhykov}
\affiliation{Physikalisch-Technische Bundesanstalt, Bundesallee 100, D-38116 Braunschweig, Germany}
\affiliation{Institut für Mathematische Physik, Technische Universität Braunschweig, Mendelssohnstrasse 3, D-38106 Braunschweig, Germany}
\affiliation{Laboratory for Emerging Nanometrology Braunschweig, Langer Kamp 6a/b, D-38106 Braunschweig, Germany}

\date{\today}

\begin{abstract}
The twisted light modes used in modern atomic physics experiments can be contaminated by small admixtures of plane wave radiation. Although these admixtures hardly reveal themselves in the beam intensity profile, they may seriously affect the outcome of high precision spectroscopy measurements. In the present study we propose a method for diagnosing such a plane wave contamination, which is based on the analysis of the magnetic sublevel population of atoms or ions interacting with the ``twisted + plane wave'' radiation. In order to theoretically investigate the sublevel populations, we solve the Liouville-von Neumann equation for the time evolution of atomic density matrix. The proposed method is illustrated for the electric dipole $5s \, {}^{2}\mathrm{S}_{1/2} \, – \, 5p \, {}^{2}\mathrm{P}_{3/2}$ transition in Rb induced by (linearly, radially, or azimuthally polarized) vortex light with just a small contamination. We find that even tiny admixtures of plane wave radiation can lead to remarkable variations in the populations of the ground-state magnetic sublevels. This opens up new opportunities for diagnostics of twisted light in atomic spectroscopy experiments.
\end{abstract}

\maketitle
\section{\label{sec:Introduction}Introduction}

\ptitle{Background twisted light} For more than 30 years, twisted light has attracted considerable interest in many areas of modern physics. In contrast to conventional plane waves, such beams exhibit a highly inhomogeneous intensity profile, a complex polarization texture, and a phase singularity \cite{PadgettPRSA2014,BliokhPR2015,RubinszteinDunlopJO2017}. Twisted beams have found their application in optical traps \cite{LiOL2012,KennedyOC2014} and tweezers \cite{FriesePRA1996,PadgettNP2011}, classical and quantum communication \cite{MirhosseiniNJP2015,NdaganoJLT2018}, super-resolution optical sensing \cite{DrechslerPRL2021,ZengJO2023} and imaging \cite{FuerhapterOE2005,WangSR2015}, as well as atomic magnetometers \cite{QiuPR2021,CastellucciPRL2021}. Moreover, they were recently used for high precision spectroscopy of trapped ions. In particular, Laguerre-Gaussian (LG) beams were employed to coherently excite clock transitions in single Ca${}^{+}$ \cite{SchmiegelowNatCom2016} and Yb${}^{+}$ ions \cite{LangePRL2022}. The interpretation of such experiments, however, can be complicated by incomplete knowledge of the radiation composition. For the analysis of Ca${}^{+}$ experiment, for instance, Afanasev \textit{et al.} \cite{AfanasevNJoP2018} inferred that the initially assumed circularly polarized light was slightly elliptically polarized. Furthermore, a radially polarized beam produced by a vortex retarder was suspected to be contaminated by a small amount of plane wave admixture in the Yb${}^{+}$ experiment of Lange \textit{et al.} \cite{LangePRL2022}. Such impurity of incident light is one of the major challenges to tackle in high precision spectroscopy experiments with twisted radiation.

A conventional method to obtain information about the mode composition of radiation is to analyze its intensity profile. This approach has a natural limitation in the domain of relatively small admixtures to the leading mode, since they do not cause noticeable changes in intensity distribution. Although tiny impurities are ``invisible'' for the conventional method, they may significantly affect the population dynamics of a target atom as usually observed in the form of Rabi oscillations. In this contribution, we discuss the effect of admixture of plane wave radiation to the leading twisted mode and propose an approach to investigate this admixture, which shows its full potential in the case of tiny impurities. Our approach is based on the analysis of the populations of the ground-state magnetic sublevels of an atom interacting with laser radiation. These populations can be measured, for example, by state-dependent fluorescence \cite{SchmiegelowNatCom2016,BlattEJP1988}.

The present work is organized as follows. In Sec.~\ref{subsec:TheoryLight} we briefly recall the basic formulas needed to describe the incident radiation and define the geometry for the light-atom coupling. \textcolor{black}{This coupling is described by transition matrix elements whose evaluation is discussed in Sec.~\ref{subsec:MatrixElement}. In order to analyze the time evolution of atomic populations, in Sec.~\ref{subsec:DensityMatrix} we lay down the density matrix formalism based on the Liouville-von Neumann equation. Substituting the transition matrix elements into the Liouville-von Neumann equation, we compute elements of the density matrix at each instant of time.} For analysis and guidance of experimental studies, however, it is more convenient to describe the system in terms of the so-called statistical tensors. These tensors are related to the population of atomic sublevels and characterize the orientation of the system, as shown in Sec.~\ref{subsec:Orientation}. The general theory is applied to the specific case of the $5s \, {}^{2}\mathrm{S}_{1/2} \, – \, 5p \, {}^{2}\mathrm{P}_{3/2}$ transition in Rb induced by a Bessel beam with a small admixture of a plane wave. The calculations presented in Sec.~\ref{sec:Results} indicate that even tiny admixtures can significantly affect the population of the $M_g = \pm 1/2$ ground-state sublevels. Moreover, we show how the impurity effects can be controlled by applying an external magnetic field. Finally, a summary of our results and an outlook are given in Sec.~\ref{sec:Summary}.

\section{\label{sec:Theory}Theory}

\subsection{\label{subsec:TheoryLight}Twisted light modes}

Modern experiments on the interaction of trapped atoms or ions with twisted radiation usually employ LG modes. Theoretical analysis of the coupling between these modes and atoms is a rather complicated task which can be simplified by approximating LG radiation with a Bessel beam. Such an approximation is well justified when an atom is located in the vicinity of the beam center \cite{LangePRL2022,PeshkovAdP2023}. Similarly, the radiation in the center of a Gaussian LG${}_{00}$ mode can be approximated by a plane wave when the spatial mode extent is large compared to the atomic sample. The vector potentials for both Bessel and plane waves will be introduced below.

\subsubsection{\label{subsubsec:VectorPotential}Photon vector potential}

\ptitle{Intro to plane waves and twisted light}

Since the interaction of atoms with twisted and plane wave radiation has already been widely discussed in the literature \cite{DavisJO2013,BarnettJO2013,RodriguesJPB2016,NovikovaOL2016,QuinteiroPRL2017,BabikerJO2019,BerakdarNatPhot2020}, we present here only a few basic formulas needed for our theoretical analysis. We start with the vector potential for a plane wave which may be written in the Coulomb gauge as

\begin{equation}
    \boldsymbol{A}^\mathrm{(pl)}_\lambda (\boldsymbol{r}) = A_0 \, \boldsymbol{e}_{\boldsymbol{k} \lambda} e^{i \boldsymbol{k} \cdot \boldsymbol{r}} \, ,
\label{eq:VP_PW}
\end{equation}

\noindent where $\boldsymbol{k}$ and $\boldsymbol{e}_{\boldsymbol{k} \lambda}$ are the photon wave and polarization vectors, $\omega=k c$ is its frequency, $\lambda = \pm 1$ denotes the helicity, and $A_0$ is the amplitude to be specified later. A Bessel beam is in turn characterized by the vector potential

\begin{equation}
    \boldsymbol{A}^\mathrm{(tw)}_{m_\gamma , \lambda} (\boldsymbol{r}) = A_0 \int a_{\varkappa m_\gamma}(\boldsymbol{k}_\perp) \, \boldsymbol{e}_{\boldsymbol{k} \lambda} e^{i \boldsymbol{k} \cdot \boldsymbol{r}} \, \frac{\mathrm{d}^2 \boldsymbol{k}_\perp}{(2 \pi)^2} \, ,
\label{eq:VP_TW}
\end{equation}

\noindent with the weight function

\begin{equation}
    a_{\varkappa m_\gamma}(\boldsymbol{k}_\perp) = \frac{2 \pi}{\varkappa} (-i)^{m_\gamma} e^{i m_\gamma \phi_k} \delta (k_\perp - \varkappa) \, .
\label{eq:Weight_Factor}
\end{equation}

\noindent The latter vector potential describes a beam with the amplitude $A_0$, the helicity $\lambda$, the longitudinal $k_z$ and transverse $\varkappa$ components of the linear momentum, as well as the projection $m_\gamma$ of the total angular momentum onto the light propagation direction \cite{SerboUsp2018,SchulzPhysRevA2020}. It follows from Eqs.~(\ref{eq:VP_PW})-(\ref{eq:Weight_Factor}) that the Bessel beam can be seen as a coherent superposition of plane waves whose wave vectors $\boldsymbol{k}$ are uniformly distributed upon the surface of a cone with a polar opening angle $\theta_k = \mathrm{arctan} (\varkappa / k_z)$.

\begin{figure}[t]
	\centering
	\def\svgwidth{255pt}
	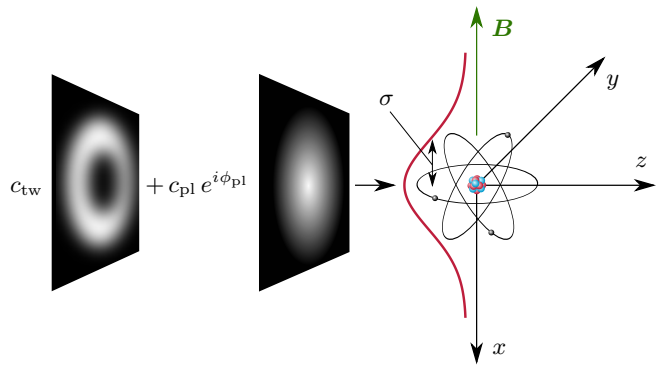
	\caption{Geometry for the excitation of a single atom by a superposition (\ref{eq:VP_Admixture}) of Bessel and plane waves. The quantization axis is chosen along the applied magnetic field, which is perpendicular to the light propagation direction. The atom is either perfectly localized in the beam center or delocalized with distribution width $\sigma$.}
	\label{fig:Geometry}
\end{figure}

In Eqs.~(\ref{eq:VP_PW})-(\ref{eq:Weight_Factor}), we introduced helicity states which are related to circularly polarized light. The other polarizations can readily be constructed from these helicity states. For example, plane waves that are linearly polarized parallel or perpendicular to a reaction plane \cite{1957_Rose}, defined by the direction of light propagation and an external magnetic field $\boldsymbol{B}$, are

\begin{subequations}
\begin{align}
    \boldsymbol{A}^\mathrm{(pl)}_x =& \, \frac{1}{\sqrt{2}} \left[ \boldsymbol{A}^\mathrm{(pl)}_{\lambda=-1} + \boldsymbol{A}^\mathrm{(pl)}_{\lambda=+1} \right] \, , \\
    \boldsymbol{A}^\mathrm{(pl)}_y =& \, \frac{i}{\sqrt{2}} \left[ \boldsymbol{A}^\mathrm{(pl)}_{\lambda=-1} - \boldsymbol{A}^\mathrm{(pl)}_{\lambda=+1} \right] \, ,
\end{align}
\label{eq:VP_PW_Par_Perp}
\end{subequations}

\noindent see Fig.~\ref{fig:Geometry} for further details. In a similar way, one can also construct linearly polarized twisted modes

\begin{subequations}
\begin{align}
    \boldsymbol{A}^\mathrm{(tw)}_x =& \, \frac{i}{\sqrt{2}} \left[ \boldsymbol{A}^\mathrm{(tw)}_{\textcolor{black}{m_{\gamma_1}} , \lambda=+1} - \boldsymbol{A}^\mathrm{(\textcolor{black}{tw})}_{\textcolor{black}{m_{\gamma_2}} , \lambda=-1} \right] \, , \\
    \boldsymbol{A}^\mathrm{(tw)}_y =& \, \frac{1}{\sqrt{2}} \left[ \boldsymbol{A}^\mathrm{(tw)}_{\textcolor{black}{m_{\gamma_1}} , \lambda=+1} + \boldsymbol{A}^\mathrm{(\textcolor{black}{tw})}_{\textcolor{black}{m_{\gamma_2}} , \lambda=-1} \right] \, ,
\end{align}
\label{eq:VP_TW_Par_Perp}
\end{subequations}

\noindent \textcolor{black}{where $m_{\gamma_1}-m_{\gamma_2}=2$} as discussed in Ref.~\cite{SchulzPhysRevA2020}. Furthermore, in contrast to plane waves, twisted radiation provides a richer choice of polarization patterns. For instance, the radially and azimuthally polarized beams read as

\begin{subequations}
\begin{align}
    \boldsymbol{A}^\mathrm{(tw)}_\mathrm{rad} =& \, -\frac{i}{\sqrt{2}} \left[ \boldsymbol{A}^\mathrm{(tw)}_{m_\gamma=0 , \lambda=+1} + \boldsymbol{A}^\mathrm{(tw)}_{m_\gamma=0 , \lambda=-1} \right] \, , \\
    \boldsymbol{A}^\mathrm{(tw)}_\mathrm{az} =& \, -\frac{1}{\sqrt{2}} \left[ \boldsymbol{A}^\mathrm{(tw)}_{m_\gamma=0 , \lambda=+1} - \boldsymbol{A}^\mathrm{(tw)}_{m_\gamma=0 , \lambda=-1} \right] \, .
\end{align}
\label{eq:VP_TW_Rad_Azim}
\end{subequations}

It can be easily seen that the vector potentials~(\ref{eq:VP_TW_Par_Perp}) and (\ref{eq:VP_TW_Rad_Azim}) correspond to linear, radial, and azimuthal polarizations in the paraxial regime where the opening angle $\theta_k$ is small. In this regime the spin and orbital angular momenta are decoupled from each other, and Eq.~(\ref{eq:VP_TW}) is simplified to:

\begin{equation}
    \boldsymbol{A}_{m_\gamma , \lambda}^{(\mathrm{tw})} (\boldsymbol{r}) \approx \boldsymbol{A}_{m_l , \lambda}^{(\mathrm{par})} (\boldsymbol{r}) = \boldsymbol{e}_\lambda (-i)^\lambda J_{m_l} (\varkappa r_\perp) e^{i m_l \phi} e^{i k_z z} \, ,
\label{eq:VP_Paraxial}
\end{equation}

\noindent where $m_l = m_\gamma - \lambda$ is the projection of the orbital angular momentum, $\boldsymbol{e}_\lambda = \boldsymbol{e}_{\boldsymbol{k}\parallel z , \lambda}$ is the polarization vector, $J_{m_l} (\varkappa r_\perp)$ stands for the Bessel function, and $r_\perp$, $\phi$, $z$ are cylindrical coordinates \cite{SerboUsp2018}. By using Eqs.~(\ref{eq:VP_TW_Par_Perp})-(\ref{eq:VP_Paraxial}), we obtain

\begin{subequations}
\begin{align}
    \boldsymbol{A}^\mathrm{(par)}_x =& \, \boldsymbol{e}_x J_{m_l} (\varkappa r_\perp) e^{i m_l \phi} e^{i k_z z} \, , \\
    \boldsymbol{A}^\mathrm{(par)}_y =& \, \boldsymbol{e}_y J_{m_l} (\varkappa r_\perp) e^{i m_l \phi} e^{i k_z z} \, , \\
    \boldsymbol{A}^\mathrm{(par)}_\mathrm{rad} =& \, \boldsymbol{e}_r J_{1} (\varkappa r_\perp) e^{i k_z z} \, , \label{eq:VP_Paraxial_Rad} \\
    \boldsymbol{A}^\mathrm{(par)}_\mathrm{az} =& \, \boldsymbol{e}_\phi J_{1} (\varkappa r_\perp) e^{i k_z z} \, , \label{eq:VP_Paraxial_Az}
\end{align}
\label{eq:VP_TW_Paraxial}
\end{subequations}

\noindent where $\boldsymbol{e}_x$, $\boldsymbol{e}_y$, $\boldsymbol{e}_r$, and $\boldsymbol{e}_\phi$ are the basis unit vectors in Cartesian and cylindrical coordinates, respectively. Strictly speaking, the solutions~(\ref{eq:VP_Paraxial_Rad}) and (\ref{eq:VP_Paraxial_Az}) do not possess a well-defined orbital angular momentum projection and therefore cannot be referred to as twisted light. Usually such fields are called vector beams \cite{RosalesJOpt2018,CastellucciPRL2021}.

\subsubsection{\label{subsec:PolarizationAdmixtures}Superposition of two modes}

\ptitle{Intro to admixtures and geometrical set-up} As already mentioned above, incident radiation may not always be produced in a pure twisted state. Instead, a target atom can be exposed to a superposition of different modes which contains the LG${}_{00}$ one as well. In order to model the impact of such an admixture, we will add a plane wave component to the twisted light

\begin{equation}
    \boldsymbol{A}^\mathrm{(mix)} = c_\mathrm{tw} \boldsymbol{A}^\mathrm{(tw)} + c_\mathrm{pl} \, e^{i \phi_\mathrm{pl}}  \boldsymbol{A}^\mathrm{(pl)} \, ,
\label{eq:VP_Admixture}
\end{equation}

\noindent where the real mixture coefficients $c_\mathrm{tw}$ and $c_\mathrm{pl}$ satisfy the normalization condition $c_\mathrm{tw}^2+c_\mathrm{pl}^2 =1$ and $\phi_\mathrm{pl}$ is the relative phase of the two light modes. In general, this relative phase can affect not only the beam intensity profile, but also its polarization texture.

Below we will discuss a method for determining the weight $c_\mathrm{pl}$ and phase $\phi_\mathrm{pl}$ of the plane wave component based on the analysis of the population dynamics of magnetic sublevels in a target atom exposed to the radiation (\ref{eq:VP_Admixture}). This requires a choice of the quantization axis of the overall system. In our study we will utilize the geometry similar to what is used in Hanle effect experiments \cite{1964_Lurio,Lehmann1964,1985_Hunter,PeikLP1994,BreschiPRA2012}, i.e., the atomic quantization axis is chosen to be along the magnetic field applied perpendicular to the light propagation direction (see Fig.~\ref{fig:Geometry}).

\textcolor{black}{\subsection{\label{subsec:MatrixElement}Evaluation of the transition matrix element}}

\ptitle{Evaluation} \textcolor{black}{Having discussed the vector potentials of Bessel and plane waves, we are ready now to examine their interaction with an atom. In particular, we will question the laser-induced transition between ground $\left| \alpha_g J_g M_g \right>$ and excited $\left| \alpha_e J_e M_e \right>$ atomic states whose properties can be traced back to the first-order matrix element}

%Solution of the Liouville-von Neumann equation (\ref{eq:Liouville_von_Neumann}) also requires a knowledge of the transition matrix element

\begin{equation}
    V_{e g} = e c \left< \alpha_e J_e M_e \Biggl| \sum\limits_q  \boldsymbol{\alpha}_q \cdot \boldsymbol{A} (\boldsymbol{r}_q) \Biggr| \alpha_g J_g M_g \right> \, ,
\label{eq:Veg}
\end{equation}

\noindent \textcolor{black}{where $J$ denotes the total angular momentum, $M$ is its projection on the atomic quantization axis, and $\alpha$ refers to all additional quantum numbers. Moreover,} $q$ runs over all electrons in a target atom and $\boldsymbol{\alpha}_q$ denotes the vector of Dirac matrices for the $q$th particle \cite{2007_Johnson}. This matrix element depends on a particular choice of the vector potential $\boldsymbol{A}$. For the superposition of twisted and plane wave radiation (\ref{eq:VP_Admixture}), we have

\begin{equation}
    V_{e g}^\mathrm{(mix)} = c_\mathrm{tw} V_{e g}^\mathrm{(tw)} + c_\mathrm{pl} \, e^{i \phi_\mathrm{pl}}  V_{e g}^\mathrm{(pl)} \, .
\label{eq:Veg_Admixture}
\end{equation}

\noindent The evaluation of the twisted $V_{e g}^\mathrm{(tw)}$ and plane-wave $V_{e g}^\mathrm{(pl)}$ matrix elements has already been discussed in detail in the literature \cite{SurzhykovPRA2015,SchulzPhysRevA2020}. For the geometry shown in Fig.~\ref{fig:Geometry}, where light propagates along the $z$-axis perpendicular to the atomic quantization axis, they read

\begin{widetext}

\begin{equation}
\begin{split}
    V_{e g}^\mathrm{(pl)} (\lambda) =& \, A_0 e c \, \sqrt{2 \pi} \, i^L (i \lambda)^p \frac{[L]^{1/2}}{[J_e]^{1/2}} \, d^L_{M_e - M_g , \lambda} (\pi/2) \left< J_g \, M_g \, L \, M_e - M_g | J_e \, M_e \right> \left< \alpha_e J_e || H_\gamma (pL) || \alpha_g J_g \right> \, ,
\end{split}
\label{eq:VegFullPW}
\end{equation}

% \end{widetext}
% \newpage
% \begin{widetext}

\begin{equation}
\begin{split}
    V_{e g}^\mathrm{(tw)} (\lambda) =& \, A_0 e c \, \sqrt{2 \pi} \sum\limits_{M} i^{L+M} (i \lambda)^p (-1)^{m_\gamma} \frac{[L]^{1/2}}{[J_e]^{1/2}} \, e^{i (m_\gamma - M) \phi_b} J_{m_\gamma - M} (\varkappa b) \, d^L_{M , \lambda} (\theta_k) \, d^L_{M_e - M_g , M} (\pi/2) \\
    &\times \left< J_g \, M_g \, L \, M_e - M_g | J_e \, M_e \right> \left< \alpha_e J_e || H_\gamma (pL) || \alpha_g J_g \right> \, ,
\end{split}
\label{eq:VegFullTW}
\end{equation}

\end{widetext}

\noindent where we have assumed, moreover, that both light field components are circularly polarized. Here, the reduced matrix element $\left< \alpha_e J_e || H_\gamma (pL) || \alpha_g J_g \right>$ for a magnetic ($p=0$) or electric ($p=1$) transition of multipolarity $L$ depends on the electronic structure of an atom and its evaluation will be discussed later. Furthermore, in Eqs.~(\ref{eq:VegFullPW}) and (\ref{eq:VegFullTW}) $d^L_{M , \lambda} (\theta_k)$ is the small Wigner $D$ function and $[J]=2J+1$.

We note that the matrix element (\ref{eq:VegFullTW}) also depends on the impact parameter $\boldsymbol{b}=(b \cos\phi_b ,b \sin\phi_b ,0)$ which specifies the position of a target atom with respect to the beam center. The introduction of this parameter is necessary since the Bessel beam exhibits an inhomogeneous intensity profile with a central dark spot at $b=0$ \cite{RosalesJOpt2018}. In contrast to Bessel beams, both the intensity and phase of plane waves do not depend on spatial position, and hence there is no need to add $\boldsymbol{b}$ to Eq.~(\ref{eq:VegFullPW}).

Similar to the discussion in Sec.~\ref{subsubsec:VectorPotential}, the matrix elements (\ref{eq:VegFullPW}) and (\ref{eq:VegFullTW}) can be used as \textit{building blocks} to investigate cases of polarization different from circular. For example, in order to analyze the interaction of an atom with linearly polarized plane or twisted waves, one \textcolor{black}{should use}

\begin{subequations}
\begin{align}
    V_{e g}^\mathrm{(pl)} (x) =& \, \frac{1}{\sqrt{2}} \left[ V_{e g}^\mathrm{(pl)} (\lambda=-1) + V_{e g}^\mathrm{(pl)} (\lambda=+1) \right] \, , \\
    V_{e g}^\mathrm{(pl)} (y) =& \, \frac{i}{\sqrt{2}} \left[ V_{e g}^\mathrm{(pl)} (\lambda=-1) - V_{e g}^\mathrm{(pl)} (\lambda=+1) \right] \, , \\
    V_{e g}^\mathrm{(tw)} (x) =& \, \frac{i}{\sqrt{2}} \left[ V_{e g}^\mathrm{(tw)} (\lambda=+1) - V_{e g}^\mathrm{(tw)} (\lambda=-1) \right] \, , \\
    V_{e g}^\mathrm{(tw)} (y) =& \, \frac{1}{\sqrt{2}} \left[ V_{e g}^\mathrm{(tw)} (\lambda=+1) + V_{e g}^\mathrm{(tw)} (\lambda=-1) \right] \, .
\end{align}
\label{eq:Veg_PW_TW_Par_Perp}
\end{subequations}

\noindent In a similar way one can construct matrix elements for the interaction with radially or azimuthally polarized vector beams.

\textcolor{black}{\subsection{\label{subsec:DensityMatrix}Density matrix formalism}}

\ptitle{Intro to density matrix} \textcolor{black}{Due to the interaction of atom with light, the populations of atomic ground and excited states can vary with time. To investigate time dependence of atomic level populations, it is practical to use the time-dependent density matrix theory \cite{2010_Budker}, where a state of the system is represented by the density operator $\hat{\rho}(t)$ satisfying the Liouville-von Neumann equation:}

\begin{equation}
    \frac{\mathrm{d}}{\mathrm{d} t}\hat{\rho}(t) = -\frac{i}{\hbar} \left[ \hat{H}(t), \hat{\rho}(t) \right] + \hat{R}(t) \, .
\label{eq:Liouville_Original}
\end{equation}

\noindent Here, $\hat{H}(t)$ is the total Hamiltonian of an atom in the presence of both the magnetic field and the incident radiation. Moreover, the operator $\hat{R}(t)$ has been introduced to take into account phenomenologically the relaxation processes, see Ref.~\cite{TannoudjiJPR1961,TremblayPRA1990} for more details.

In order to express the operator $\hat{\rho}(t)$ in matrix form, a convenient set of basis states must be chosen. In our work we use the ground $\left| \alpha_g J_g M_g \right>$ and excited $\left|\alpha_e J_e M_e \right>$ atomic states as a basis. In this basis, the matrix elements of $\hat{\rho}(t)$, also known as the density matrix, take the form:

\begin{subequations}
\begin{align}
    \rho_{g g^\prime} (t) =& \, \left< \alpha_g J_g M_g | \hat{\rho} (t) | \alpha_g J_g M_g^\prime \right> \, , \label{eq:rhoggprime}\\
    \rho_{e e^\prime} (t) =& \, \left< \alpha_e J_e M_e | \hat{\rho} (t) | \alpha_e J_e M_e^\prime \right> \, , \label{eq:rhoeeprime}\\
    \rho_{g e} (t) =& \, \left< \alpha_g J_g M_g | \hat{\rho} (t) | \alpha_e J_e M_e \right> \, , \\
    \rho_{e g} (t) =& \, \left< \alpha_e J_e M_e | \hat{\rho} (t) | \alpha_g J_g M_g \right> \, .
\end{align}
\label{eq:Density_Matrix}
\end{subequations}

\noindent Here, the notation $\rho_{g g} (t)$ and $\rho_{e e} (t)$ is used as shorthand for the probability of finding an atom in the substate $\left|\alpha_g J_g M_g \right>$ and $\left|\alpha_e J_e M_e \right>$, respectively, while $\rho_{g g^\prime} (t)$ and $\rho_{e e^\prime} (t)$ describe the coherences between different substates \cite{2012_Blum}. In the present work we will focus especially on the ground-state density matrix and investigate its dependence on the magnetic field strength for different composition of incident radiation.

From Eqs.~(\ref{eq:Liouville_Original}) and (\ref{eq:Density_Matrix}), we obtain the following set of differential equations for the density matrix elements:

\begin{widetext}

\begin{subequations}
\allowdisplaybreaks
\begin{align}
\allowdisplaybreaks
    \frac{\mathrm{d}}{\mathrm{d} t}\widetilde{\rho}_{g g^\prime} (t) =& \, - i \Omega^{\mathrm{(L)}}_g \left(M_g - M_g^\prime \right) \widetilde{\rho}_{g g^\prime} (t) - \frac{i}{2 \hbar} \left[ \sum\limits_{\widetilde{M}_e} V_{\tilde{e} g}^{*} \, \widetilde{\rho}_{\tilde{e} g^\prime} (t) -  \sum\limits_{\widetilde{M}_e} V_{\tilde{e} g^\prime} \, \widetilde{\rho}_{g \tilde{e}} (t) \right] + R_{g g^\prime} (t) \, , \\
    \frac{\mathrm{d}}{\mathrm{d} t}\widetilde{\rho}_{e e^\prime} (t) =& \, - i \Omega^{\mathrm{(L)}}_e \left(M_e - M_e^\prime \right) \widetilde{\rho}_{e e^\prime} (t) - \frac{i}{2 \hbar} \left[ \sum\limits_{\widetilde{M}_g} V_{e \tilde{g}} \, \widetilde{\rho}_{\tilde{g} e^\prime} (t) -  \sum\limits_{\widetilde{M}_g} V_{e^\prime \tilde{g}}^{*} \, \widetilde{\rho}_{e \tilde{g}} (t) \right] + R_{e e^\prime} (t) \, , \\
    \frac{\mathrm{d}}{\mathrm{d} t}\widetilde{\rho}_{g e} (t) =& \, -i \Delta \widetilde{\rho}_{g e} (t) + i \left(\Omega^{\mathrm{(L)}}_e M_e - \Omega^{\mathrm{(L)}}_g M_g \right) \widetilde{\rho}_{g e} (t) - \frac{i}{2 \hbar} \left[ \sum\limits_{\widetilde{M}_e} V_{\tilde{e} g}^{*} \, \widetilde{\rho}_{\tilde{e} e} (t) - \sum\limits_{\widetilde{M}_g} V_{e \tilde{g}}^{*} \, \widetilde{\rho}_{g \tilde{g}} (t) \right] + R_{g e} (t) \, , \\
    \frac{\mathrm{d}}{\mathrm{d} t}\widetilde{\rho}_{e g} (t) =& \, i \Delta \widetilde{\rho}_{e g} (t) - i \left(\Omega^{\mathrm{(L)}}_e M_e - \Omega^{\mathrm{(L)}}_g M_g \right) \widetilde{\rho}_{e g} (t) - \frac{i}{2 \hbar} \left[ \sum\limits_{\widetilde{M}_g} V_{e \tilde{g}} \, \widetilde{\rho}_{\tilde{g} g} (t) - \sum\limits_{\widetilde{M}_e} V_{\tilde{e} g} \, \widetilde{\rho}_{e \tilde{e}} (t) \right] + R_{e g} (t) \, .
\end{align}
\label{eq:Liouville_von_Neumann}
\end{subequations}

\end{widetext}

\noindent Here we have made the substitutions

\begin{subequations}
\begin{align}
    \widetilde{\rho}_{g g^\prime} (t) =& \, \rho_{g g^\prime} (t) \, , \\
    \widetilde{\rho}_{e e^\prime} (t) =& \, \rho_{e e^\prime} (t) \, , \\
    \widetilde{\rho}_{g e} (t) =& \, \rho_{g e} (t) e^{-i \omega t} \, , \\
    \widetilde{\rho}_{e g} (t) =& \, \rho_{e g} (t) e^{i \omega t} \, ,
\end{align}
\label{eq:Substitution}
\end{subequations}

\noindent and employed the rotating-wave approximation which consists in neglecting the fast-oscillating terms proportional to $e^{\pm 2i \omega t}$ \cite{2020_Wense,1984_Stenholm}. Furthermore, $\Omega^{\mathrm{(L)}} = g_J \mu_B B / \hbar$ is the Larmor frequency, and $\Delta = \omega - \omega_0$ is the frequency detuning of the radiation from the atomic resonance at $\omega_0$.

In Eqs.~(\ref{eq:Liouville_von_Neumann}), the terms $R (t)$ account phenomenologically for the relaxation of an atom due to spontaneous emission. To derive these terms, we follow the procedure discussed in Ref.~\cite{1997_Renzoni,2010_Budker} and find

\begin{subequations}
\allowdisplaybreaks
\begin{align}
\allowdisplaybreaks
\begin{split}
    R_{g g^\prime} (t) =& \; \Gamma \sum\limits_{M_e, M_e^\prime, M} \left< J_g \, M_g \, L \, M | J_e \, M_e \right> \\
    &\, \quad \; \, \, \widetilde{\rho}_{e e^\prime} (t) \left< J_g \, M_g^\prime \, L \, M | J_e \, M_e^\prime \right> \, ,
\end{split} \\
    R_{e e^\prime} (t) =& \, - \Gamma \widetilde{\rho}_{e e^\prime} (t) \, , \\
    R_{g e} (t) =& \, - \frac{\Gamma}{2} \widetilde{\rho}_{g e} (t) \, , \\
    R_{e g} (t) =& \, - \frac{\Gamma}{2} \widetilde{\rho}_{e g} (t) \, ,
\end{align}
\label{eq:RTerms}
\end{subequations}

\noindent with $\Gamma$ being the decay rate of  excited state. In obtaining Eqs.~(\ref{eq:RTerms}), we have assumed that only one dominant channel with multipolarity $L$ contributes to the decay.
\\\\
\subsection{\label{subsec:Orientation}Statistical tensors of atomic states}

\ptitle{Orientation} Solving Eqs.~(\ref{eq:Liouville_von_Neumann}) numerically, we find the atomic density matrix at each particular moment in time. In order to visualize the results and simplify the discussion, it is more convenient to describe the population of atomic sublevels in terms of the statistical tensors \cite{2000_Balashov} that are linear combinations of the density matrix elements

\begin{equation}
\begin{split}
    \rho_{kq} (\alpha J ; t) =& \, \sum\limits_{M \, M^\prime} (-1)^{J-M^\prime} \left< J \, M \, J \, -M^\prime | k \, q \right> \\
    &\times \left< \alpha J M | \hat{\rho} (t) | \alpha J M^\prime \right> \, .
\end{split}
\end{equation}

\noindent These tensors have well-defined symmetry properties since they transform like the spherical harmonics of rank $k$ under a rotation of the coordinates. In atomic physics, $\rho_{kq}$ are usually normalized as

\begin{equation}
    \mathcal{A}_{kq} (\alpha J ; t) = \frac{\rho_{kq} (\alpha J ; t)}{\rho_{00} (\alpha J ; t)} \,
\end{equation}

\noindent to produce the so-called alignment and orientation parameters. These parameters characterize the relative population of magnetic sublevels $\left| \alpha J M \right>$ and coherence between them. If all magnetic sublevels are equally populated, the atom is unpolarized and the only nonzero parameter is $\mathcal{A}_{00}=1$. In contrast, unequal substate populations lead to at least one non-vanishing parameter $\mathcal{A}_{kq}$ with $k > 0$. If the system is characterized by parameters $\mathcal{A}_{kq}$ of even rank $k$, it is said to be aligned, while the system is called oriented if at least one odd rank parameter $\mathcal{A}_{kq}$ is nonzero \cite{2000_Balashov}.

In what follows we will investigate the population of the $5s \, {}^{2}\mathrm{S}_{1/2}$ state, which can be described by only three nontrivial parameters $\mathcal{A}_{1q}$ with $q=0, \pm 1$. Here, $\mathcal{A}_{10}$ describes the difference in the population of magnetic sublevels

\begin{equation}
    \mathcal{A}_{10} (t) = \frac{\rho_{+1/2} (t) - \rho_{-1/2} (t)}{\rho_{+1/2} (t) + \rho_{-1/2} (t)} \, ,
    \label{eq:OrientationParameter}
\end{equation}

\noindent with

\begin{equation}
    \rho_{M_g} (t) = \left< 5s \, {}^{2}\mathrm{S}_{1/2} \, M_g | \, \hat{\rho} (t) \, | 5s \, {}^{2}\mathrm{S}_{1/2} \, M_g \right> \, ,
\end{equation}

\noindent and hence characterizes the orientation of the $5s \, {}^{2}\mathrm{S}_{1/2}$ state, while parameters $\mathcal{A}_{1\pm 1}$ reflect coherences between different substates.

In the next section, we will analyze the dependence of the orientation parameters $\mathcal{A}_{1q}$ on the external magnetic field strength for different mixtures of incident radiation to determine the weight and phase of the plane wave admixture.

\section{\label{sec:Results}Results and Discussion}

\begin{figure}[t]
	\raggedright
	\def\svgwidth{225pt}
	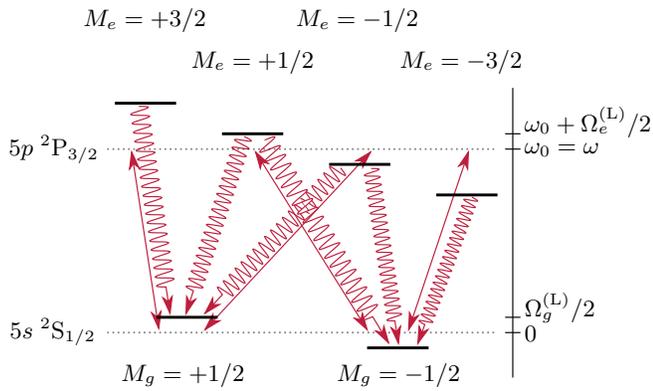
	\caption{The transition scheme for the $5 s \, {}^{2}\mathrm{S}_{1/2} - 5 p \, {}^{2}\mathrm{P}_{3/2}$ excitation of rubidium, along with the Zeeman splitting of the magnetic sublevels. The arrows represent the transition as induced by $y$-polarized incident light (\ref{eq:VP_Admixture}) with $m_l = +1$, and the wavy lines represent spontaneous decay.}
	\label{fig:System}
\end{figure}

In the previous section we have derived the Liouville-von Neumann equation~(\ref{eq:Liouville_von_Neumann}) which allows us to investigate the time-dependent interaction of an atom with a beam propagating along the $z$-axis in the presence of a magnetic field ($\boldsymbol{B} \perp \boldsymbol{e}_z$). Below we will use this theory to explore the interaction of Rb atom, initially prepared in the unpolarized $5s \, {}^{2}\mathrm{S}_{1/2}$ state, with the superposition (\ref{eq:VP_Admixture}) of twisted and plane waves. Both modes are supposed to drive the $5s \, {}^{2}\mathrm{S}_{1/2} - 5p \, {}^{2}\mathrm{P}_{3/2}$ electric dipole (E1) transition of frequency $\omega_0 = 2 \pi \times 384$ THz, see Fig.~\ref{fig:System}.

The numerical solution of Eqs.~(\ref{eq:Liouville_von_Neumann}) requires further information about the incident radiation and the target atom. In particular, we need to know the spontaneous decay rate $\Gamma$ and the reduced matrix element $\left< 5p \, {}^{2}\mathrm{P}_{3/2} || H_\gamma (E1) || 5s \, {}^{2}\mathrm{S}_{1/2} \right>$ which enter into Eqs.~(\ref{eq:Veg_Admixture})-(\ref{eq:VegFullTW}) and (\ref{eq:RTerms}). Their values were obtained using the package JAC which is developed to calculate energies and transition probabilities in many-electron atoms \cite{2019_Fritzsche}. Moreover, the light amplitude $A_0 = 2.54 \times 10^{-12}$ and the opening angle $\theta_k = 2.49$ deg are chosen so that the plane (\ref{eq:VP_PW}) and twisted (\ref{eq:VP_TW}) waves reproduce the LG${}_{00}$ and LG${}_{01}$ modes with the total power 4 $\mu$W and the waist 7 $\mu$m in the vicinity of the beam center. Finally, we assume that the detuning of the light from the atomic resonance in the absence of a magnetic field is zero, $\Delta = 0$.

\subsection{\label{subsec:LocalizedAtoms}Localized atom in the absence of magnetic field}

\ptitle{Results for localized atoms}

\begin{figure}[t]
	\centering
	\includegraphics[scale=0.445]{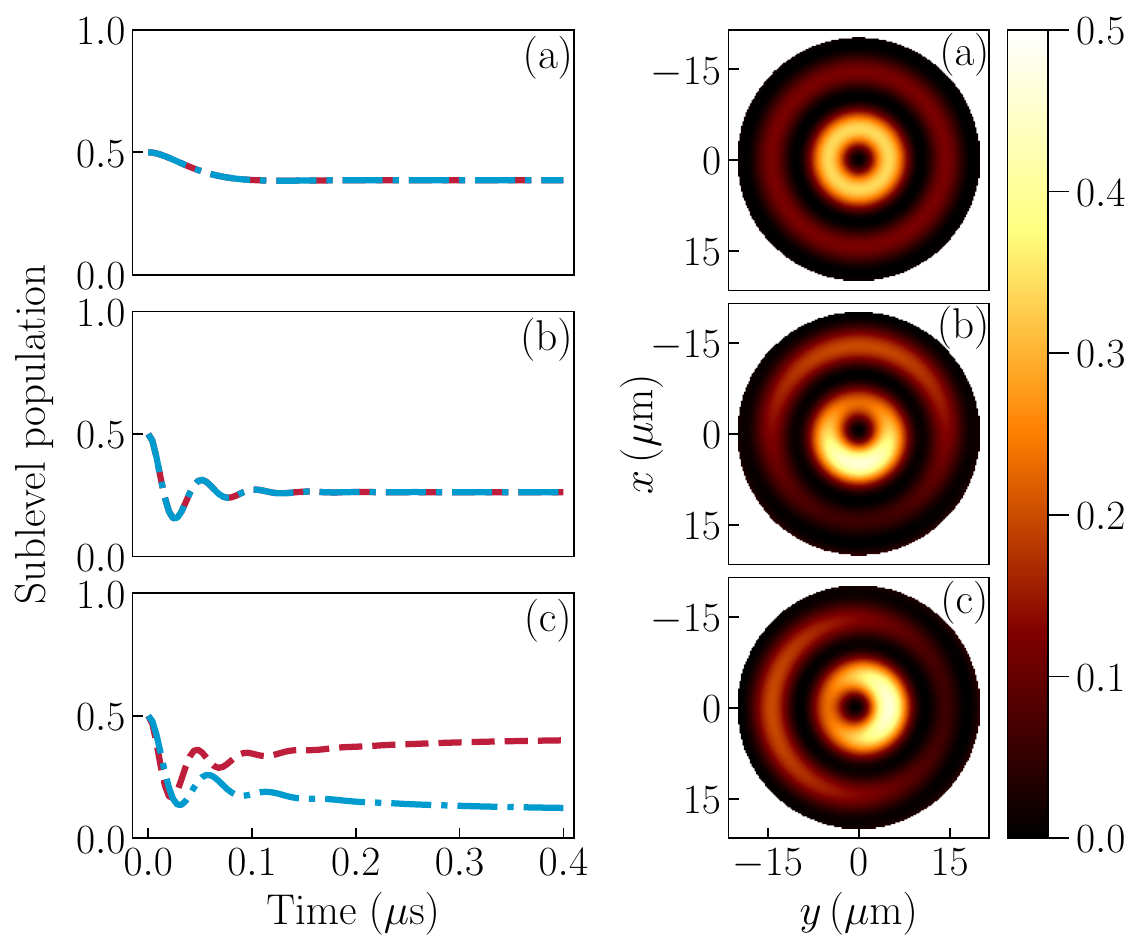}
	\caption{Left: Populations of the $M_g=-1/2$ (dashed line) and $M_g=+1/2$ (dash-dotted line) magnetic sublevels of the $5 s \, {}^{2}\mathrm{S}_{1/2}$ ground state of rubidium well localized on the vortex line, $b=0$, as a function of time for zero magnetic field, $B=0$. The incident light (\ref{eq:VP_Admixture}) is assumed to be a superposition of $y$-polarized plane and Bessel waves, where the latter carries the orbital angular momentum projection $m_l=+1$ and has the opening angle $\theta_k= 2.49$ deg. Results are presented for (a) pure Bessel beam, $c_\mathrm{pl}=0$, and superpositions with (b) $c_\mathrm{pl}=0.1$, $\phi_\mathrm{pl}=0$ and (c) $c_\mathrm{pl}=0.1$, $\phi_\mathrm{pl}=90$ deg. Right: Transverse intensity profiles of these beams in units of $A_0^2$.}
	\label{fig:TimeDependency}
\end{figure}

As seen from Eq.~(\ref{eq:VegFullTW}), we have to agree about the value of the impact parameter $\boldsymbol{b}$ to find a solution of the Liouville-von Neumann equation~(\ref{eq:Liouville_von_Neumann}). In this section, we consider an idealized scenario in which the atom is well localized on the vortex line at $b=0$. For this scenario, the left column of Fig.~\ref{fig:TimeDependency} displays the time evolution of the populations $\rho_{-1/2} (t)$ and $\rho_{+1/2} (t)$ of the ground-state magnetic sublevels $M_g = \pm 1/2$. The calculations were made in the limit of vanishing magnetic field, $B=0$, for different superpositions (\ref{eq:VP_Admixture}) of twisted and plane waves. Moreover, we assumed that both components of light, $\boldsymbol{A}^\mathrm{(tw)}$ with $m_l=+1$ and $\boldsymbol{A}^\mathrm{(pl)}$, are linearly polarized along the $y$-axis, see Fig.~\ref{fig:Geometry}. As illustrated in Fig.~\ref{fig:TimeDependency}, the population dynamics is very sensitive to the composition of incident light. For example, if the atom interacts with pure Bessel radiation, $c_\mathrm{pl}=0$, both $M_g = \pm 1/2$ sublevels are always equally populated, $\rho_{-1/2} (t) = \rho_{+1/2} (t)$. This resembles the outcome of photoexcitation by linearly polarized plane waves which is known to produce no orientation of the target along an axis normal to both light propagation and polarization directions. A similar result is obtained for the superposition of twisted and plane waves with relative phase $\phi_\mathrm{pl}=0$, see the left panel of Fig.~\ref{fig:TimeDependency}~(b). \textcolor{black}{In contrast, qualitatively different behavior can be observed when the Bessel and plane wave components are phase-shifted with respect to each other. This effect is most pronounced for the case $\phi_\mathrm{pl}=90$ deg which is displayed in Fig.~\ref{fig:TimeDependency}~(c). As seen from the figure, the populations of the $M_g = \pm 1/2$ sublevels gradually diverge from each other as time progresses and reach the values $\rho_{-1/2} = 0.40$ and $\rho_{+1/2} = 0.12$ for the steady state.} This result clearly indicates that the admixture of a plane wave to a Bessel wave can lead to significant orientation of the $5 s \, {}^{2}\mathrm{S}_{1/2}$ ground state, even though both components of the beam are linearly polarized.

\begin{figure}[t]
	\centering
	\includegraphics[scale=0.535]{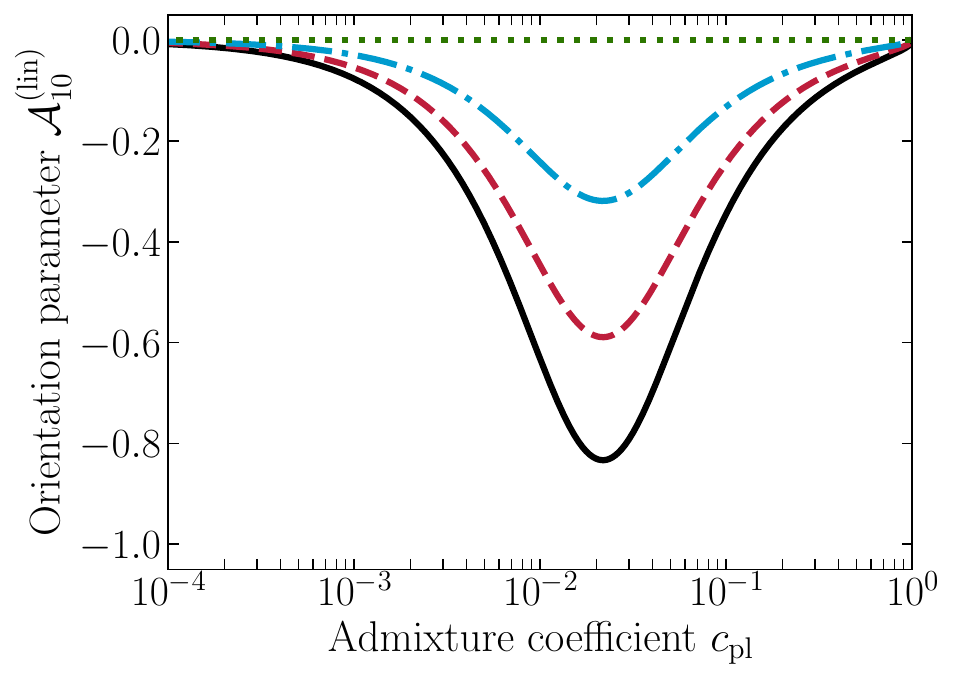}
	\caption{Orientation parameter (\ref{eq:TwoStepA10}) of the $5 s \, {}^{2}\mathrm{S}_{1/2}$ state of Rb as a function of the weight $c_\mathrm{pl}$ of the plane wave component obtained from the perturbative analysis of photoexcitation-and-decay for $B=0$. Calculations were performed for $\phi_\mathrm{pl}=90$ deg (solid line), $\phi_\mathrm{pl}=45$ deg (dashed line), $\phi_\mathrm{pl}=22.5$ deg (dash-dotted line), and $\phi_\mathrm{pl}=0$ (dotted line). All other parameters are the same as in Fig.~\ref{fig:TimeDependency}.}
	\label{fig:Analytic}
\end{figure}

To explain the qualitatively different behavior of the sublevel populations $\rho_{\pm 1/2} (t)$, in the right column of Fig.~\ref{fig:TimeDependency} we compare three different transverse intensity profiles of the incident beams. As seen from Fig.~\ref{fig:TimeDependency} (a), the intensity profile of a pure Bessel beam has the well-known annular structure which is axially symmetric with respect to the vortex line crossing the transverse plane at $x=y=0$. This symmetry is broken by an admixture of plane wave radiation. \textcolor{black}{The symmetry breaking can be easily understood if we consider the absolute value squared of the vector potential describing the $y$-polarized superposition:}

\textcolor{black}{
\begin{equation}
\begin{split}
    &\, \left| \boldsymbol{A}^\mathrm{(mix)} (r,\phi,z=0) \right|^2 \\
    \simeq& \, \left| \boldsymbol{A}^\mathrm{(tw)}(r) e^{i m_l \phi} + \boldsymbol{A}^\mathrm{(pl)}(r) e^{i \phi_\mathrm{pl}} \right|^2 \\
   =& \, \left| \boldsymbol{A}^\mathrm{(tw)}(r) \right|^2 + \left| \boldsymbol{A}^\mathrm{(pl)}(r) \right|^2 \\
   &+ \, 2 \boldsymbol{A}^\mathrm{(tw)}(r) \boldsymbol{A}^\mathrm{(pl)}(r) \cos\left(m_l \phi - \phi_\mathrm{pl} \right) \, .
\end{split}
\end{equation}
}

\noindent \textcolor{black}{As seen from this equation, the interference term containing $\cos(m_l \phi-\phi_\mathrm{pl})$ depends on the azimuthal angle $\phi$ and hence violates the axial symmetry of the beam. In addition,} the resulting asymmetric intensity profile of the ``Bessel wave + plane wave'' mixture depends on the relative phase $\phi_\mathrm{pl}$. For $\phi_\mathrm{pl}=0$, the incident beam is symmetric with respect to the $x$-$z$ plane containing the quantization axis and the light propagation direction, while this is not the case when $\phi_\mathrm{pl}=90$ deg. This difference in intensity profiles is reflected in the qualitatively different behavior of the sublevel populations. Indeed, it follows from symmetry considerations that the statistical tensor $\mathcal{A}_{10}$ vanishes for a system ``atom + light'' with the quantization axis in the plane of symmetry \cite{2000_Balashov}. This is the case for the pure Bessel beam (\ref{eq:VP_TW_Par_Perp}) and the superposition (\ref{eq:VP_Admixture}) with $\phi_\mathrm{pl}=0$, see panels (a) and (b) of Fig.~\ref{fig:TimeDependency}. In contrast, a system with broken symmetry with respect to the $x$-$z$ plane is characterized by $\mathcal{A}_{10} \neq 0$, implying $\rho_{-1/2} (t) \neq \rho_{+1/2} (t)$, see Eq.~(\ref{eq:OrientationParameter}). This is the case for the superposition (\ref{eq:VP_Admixture}) with $\phi_\mathrm{pl} = 90$ deg, displayed in panel (c) of Fig.~\ref{fig:TimeDependency}.

Apart from analysis of the beam intensity profiles, yet another approach can be used to understand---at least qualitatively---the behavior of the magnetic sublevel populations $\rho_{\pm 1/2} (t)$. This approach is based on the perturbative analysis of the excitation-and-decay of a target atom interacting with incoming twisted light. It is very close to that used for description of the resonant elastic scattering discussed in details in Refs.~\cite{1999_Roy,SerboAdP2021}, and employs second-order perturbation theory. For brevity, we will not repeat the calculation steps here and just present the perturbative prediction for the orientation parameter of the $5 s \, {}^{2}\mathrm{S}_{1/2}$ state:

\begin{equation}
    \mathcal{A}_{10}^\mathrm{(lin)}=-\frac{10 c_\mathrm{pl}\sqrt{1-c_\mathrm{pl}^2}\sin(\theta_k)\sin(\phi_\mathrm{pl})}{12 c_\mathrm{pl}^2 + 3 ( 1-c_\mathrm{pl}^2 ) \sin^2(\theta_k)} \, .
    \label{eq:TwoStepA10}
\end{equation}

\begin{figure}[t]
	\centering
	\includegraphics[scale=0.55]{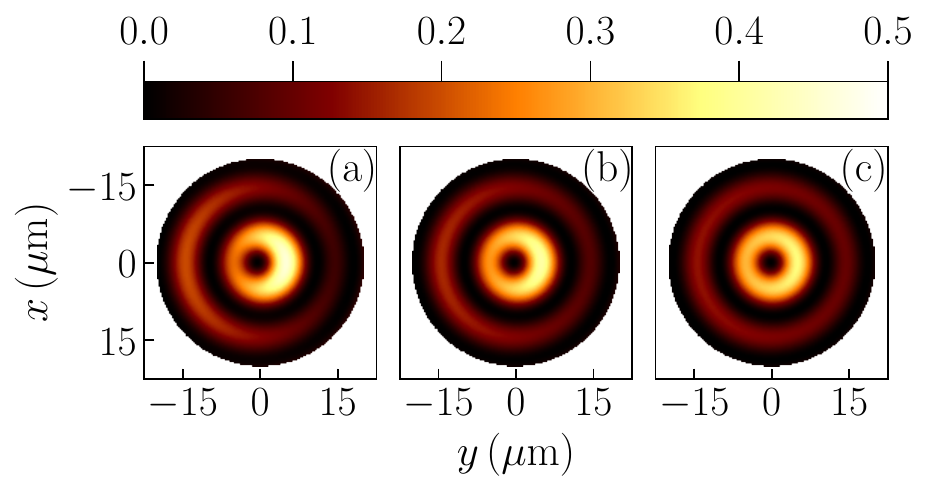}
	\caption{Same as the right column of Fig.~\ref{fig:TimeDependency}, but for $\phi_\mathrm{pl}=90$ deg and (a) $c_\mathrm{pl}=0.1$, (b) $c_\mathrm{pl}=0.06$, (c) $c_\mathrm{pl}=0.03$.}
	\label{fig:Intensity}
\end{figure}

\noindent This result is obtained for the vector potential (\ref{eq:VP_Admixture}), and hence depends on the weight $c_\mathrm{pl}$ and phase $\phi_\mathrm{pl}$ of the plane wave component, as well as on the opening angle $\theta_k$ of the Bessel beam. \textcolor{black}{Fig.~\ref{fig:Analytic} shows the predictions of Eq.~(\ref{eq:TwoStepA10}) as a function of the weight $c_\mathrm{pl}$ for $\theta_k = 2.49$ deg and different relative phases $\phi_\mathrm{pl}$.} As seen from the figure, the orientation parameter $\mathcal{A}_{10}^\mathrm{(lin)}$ vanishes in the limit of a pure plane wave, $c_\mathrm{pl}=1$, and a pure Bessel beam, $c_\mathrm{pl}=0$. This thus confirms the results of the time-dependent density matrix calculations. Moreover, perturbation theory predicts a strong dependence of $\mathcal{A}_{10}^\mathrm{(lin)}$ on the relative phase $\phi_\mathrm{pl}$ between the Bessel and plane wave components. For example, when $\phi_\mathrm{pl}=0$, there is no orientation of the ground state, regardless of the weight $c_\mathrm{pl}$. This agrees with the conclusion based on the analysis of intensity profiles. In addition, remarkable orientation of the ground state can be observed for $\phi_\mathrm{pl} \neq 0$ and weight coefficients $0.001 \leq c_\mathrm{pl} \leq 0.1$. For instance, for $\phi_\mathrm{pl}=90$ deg and $c_\mathrm{pl}=0.1$ naive perturbation theory predicts $\mathcal{A}_{10}^\mathrm{(lin)}=-0.34$, in qualitative agreement with the result obtained by solving the Liouville-von Neumann equation, $\mathcal{A}_{10}^\mathrm{(lin)}=-0.56$. \textcolor{black}{We also note from Eq.~(\ref{eq:TwoStepA10}) that the peak position $c^\mathrm{(peak)}_\mathrm{pl}$ of the orientation parameter $A_{10}^\mathrm{(lin)}$ depends on the opening angle $\theta_k$:}

\textcolor{black}{
\begin{equation}
    c^\mathrm{(peak)}_\mathrm{pl}=\frac{\sin(\theta_k)}{\sqrt{4+\sin^2(\theta_k)}} \, . 
\end{equation}
}

\noindent \textcolor{black}{In particular, for $\theta_k = 2.49$ deg the peak is located at $c^\mathrm{(peak)}_\mathrm{pl}=0.022$, see Fig.~\ref{fig:Analytic}.}

\begin{figure}[t]
	\centering
	\includegraphics[scale=0.54]{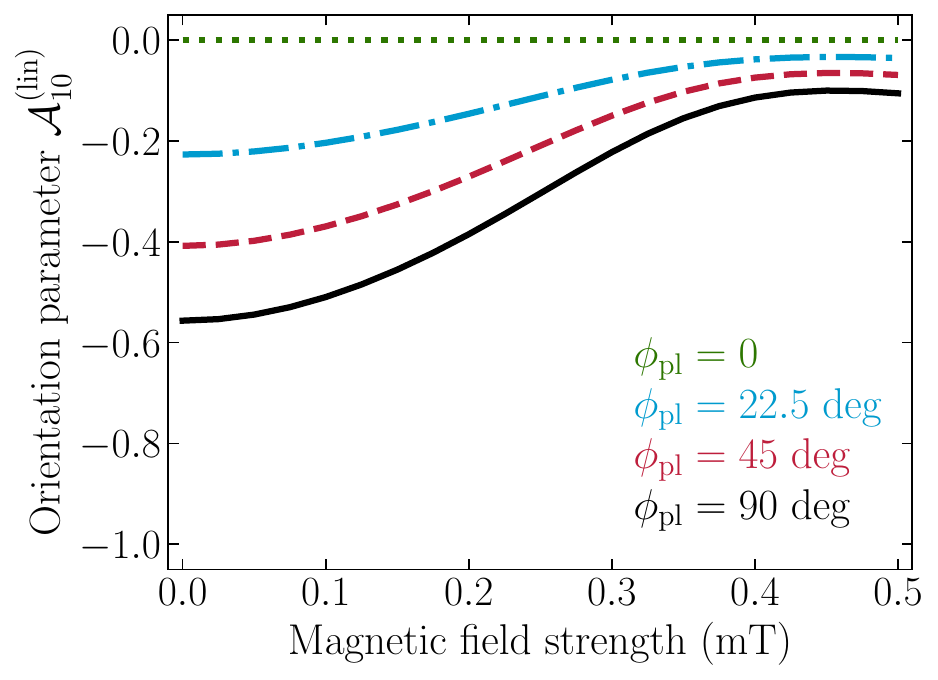}
        \caption{Orientation parameter $\mathcal{A}_{10}^\mathrm{(lin)}$ of the $5 s \, {}^{2}\mathrm{S}_{1/2}$ state of Rb as a function of the magnetic field strength for fixed weight $c_\mathrm{pl}=0.1$ and different relative phases of the plane wave component: $\phi_\mathrm{pl}=90$ deg (solid line), $\phi_\mathrm{pl}=45$ deg (dashed line), $\phi_\mathrm{pl}=22.5$ deg (dash-dotted line), and $\phi_\mathrm{pl}=0$ (dotted line). All other parameters are the same as in Fig.~\ref{fig:TimeDependency}.}
	\label{fig:FixedcPw0dot1}
\end{figure}

Both Eq.~(\ref{eq:TwoStepA10}), based on the perturbation theory approach, and the more accurate density matrix predictions of Eqs.~(\ref{eq:Liouville_von_Neumann}) show that the orientation parameter of the $5 s \, {}^{2}\mathrm{S}_{1/2}$ ground state is very sensitive to the phase $\phi_\mathrm{pl}$ and the weight $c_\mathrm{pl}$ of the plane wave component. This sensitivity is most pronounced for rather small weights in the range from $0.001$ to $0.1$. For such tiny parameters $c_\mathrm{pl}$, it is very difficult to infer any composition of the incident beam (\ref{eq:VP_Admixture}) from the intensity profile. For example, the superpositions with $c_\mathrm{pl}=0.1$, $c_\mathrm{pl}=0.06$, and $c_\mathrm{pl}=0.03$ exhibit very similar intensity profiles, as displayed in Fig.~\ref{fig:Intensity}, while the corresponding orientation parameters $\mathcal{A}_{10}^\mathrm{(lin)}=-0.34$, $\mathcal{A}_{10}^\mathrm{(lin)}=-0.53$, and $\mathcal{A}_{10}^\mathrm{(lin)}=-0.79$ are clearly distinguishable and relatively easy to observe in modern experiments.

\subsection{\label{subsec:MagneticFieldDependence}Magnetic-field dependence}

In the previous section we have shown that the orientation of the $5 s \, {}^{2}\mathrm{S}_{1/2}$ ground state of the target atom is very sensitive to the weight and relative phase of the plane wave admixture to the dominant twisted mode. We argue, therefore, that measurements of the atomic orientation can be used to study the beam composition. In order to make the proposed diagnostic method even more accessible, it is convenient to introduce one more physical parameter whose variation would affect $\mathcal{A}_{10}$. The applied magnetic field strength $B$ may be such a parameter. As we have already mentioned, the atomic quantization axis is chosen to be along $\boldsymbol{B}$, which is perpendicular to both light propagation and polarization directions.

\begin{figure}[t]
	\centering
	\includegraphics[scale=0.54]{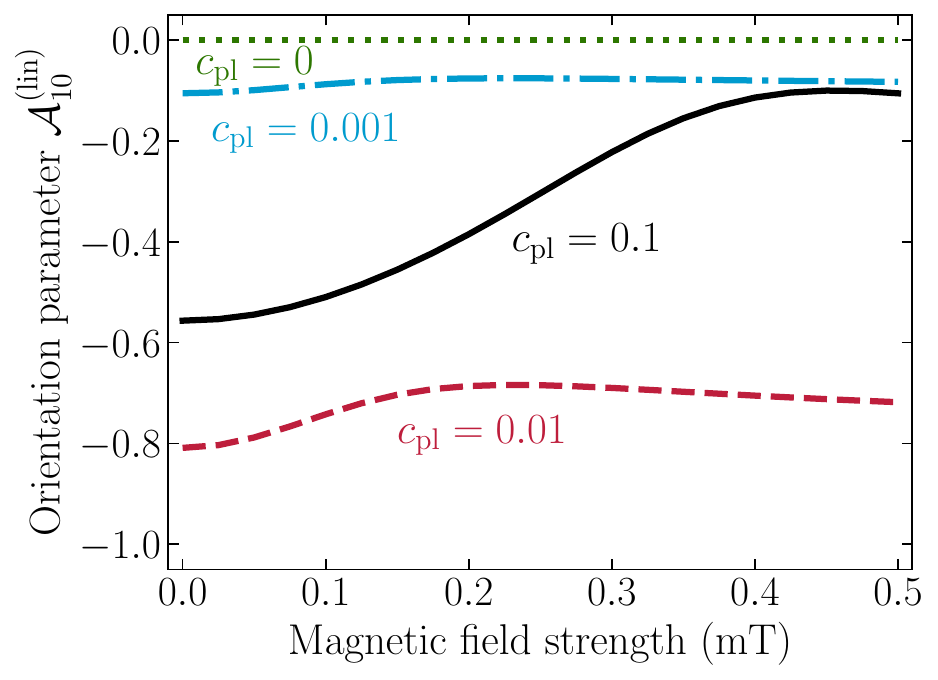}
        \caption{Same as Fig. \ref{fig:FixedcPw0dot1}, but for fixed phase $\phi_\mathrm{pl} = 90$ deg and different weights of the plane wave component: $c_\mathrm{pl}=0.1$ (solid line), $c_\mathrm{pl}=0.01$ (dashed line), $c_\mathrm{pl}=0.001$ (dash-dotted line), and $c_\mathrm{pl}=0$ (dotted line).}
	\label{fig:FixedPhiPiHalf}
\end{figure}

The magnetic-field dependence of the steady-state sublevel population of the $5 s \, {}^{2}\mathrm{S}_{1/2}$ state, produced in the course of the interaction with the $y$-polarized ``Bessel wave + plane wave'' mixture, is shown in Fig.~\ref{fig:FixedcPw0dot1}. Here calculations were performed for fixed weight $c_\mathrm{pl}=0.1$ but different phases $\phi_\mathrm{pl}$. Moreover, the magnetic field strength lies in the range $0 \leq B \leq 0.5$ mT, so that the Zeeman splitting of the $5 p \, {}^{2}\mathrm{P}_{3/2}$ level is comparable with the natural width of the transition. As seen from the figure, the orientation parameter is very sensitive to $B$. In particular, the orientation of the ground state is most pronounced for small values of magnetic field and then decreases with $B$. For $\phi_\mathrm{pl}=90$ deg, the orientation parameter takes the value $\mathcal{A}_{10}^\mathrm{(lin)} \approx -0.56$ at $B=0$, but is reduced to $\mathcal{A}_{10}^\mathrm{(lin)} \approx -0.11$ at $B=0.5$ mT. This behavior agrees with the predictions of perturbation theory. Indeed, a formula similar to Eq.~(\ref{eq:TwoStepA10}) can be derived for a non-vanishing magnetic field which shows that $\mathcal{A}_{10}^\mathrm{(lin)}(B)$ is monotonically decreasing. This formula is rather complicated and for brevity will not be shown here.

While Fig.~\ref{fig:FixedcPw0dot1} shows $\mathcal{A}_{10}^\mathrm{(lin)} (B)$ for different relative phases $\phi_\mathrm{pl}$, Fig.~\ref{fig:FixedPhiPiHalf} illustrates how the orientation of the $5 s \, {}^{2}\mathrm{S}_{1/2}$ state varies with weight $c_\mathrm{pl}$. In this figure we find an ordering of $\mathcal{A}_{10}^\mathrm{(lin)}$ which might be seen counterintuitive at first sight. Namely, while atomic orientation vanishes in the limit of a pure Bessel beam, $c_\mathrm{pl}=0$, we obtain $\mathcal{A}_{10}^\mathrm{(lin)} \approx 0.1$ for $c_\mathrm{pl}=0.001$. Then the parameter $\mathcal{A}_{10}^\mathrm{(lin)}$ reaches a maximum absolute value when $c_\mathrm{pl}=0.01$ and decreases again for $c_\mathrm{pl}=0.1$. This behavior is not surprising since it reflects the $c_\mathrm{pl}$-dependence of $\mathcal{A}_{10}^\mathrm{(lin)}$ shown in Fig.~\ref{fig:Analytic}. We also see from Fig.~\ref{fig:FixedPhiPiHalf} that the sensitivity of the orientation parameter $\mathcal{A}_{10}^\mathrm{(lin)}(B)$ depends on the weight $c_\mathrm{pl}$. For example, while $\mathcal{A}_{10}^\mathrm{(lin)}$ varies from $-0.56$ to $-0.11$ for $c_\mathrm{pl}=0.1$, it remains almost constant for $c_\mathrm{pl}=0.001$. Such a $B$-dependence can be naturally used in experiments to analyze the weight of a plane wave admixture.

\subsection{\label{subsec:DelocalizedAtoms}Delocalized atom}

\begin{figure}[t]
	\centering
	\includegraphics[scale=0.54]{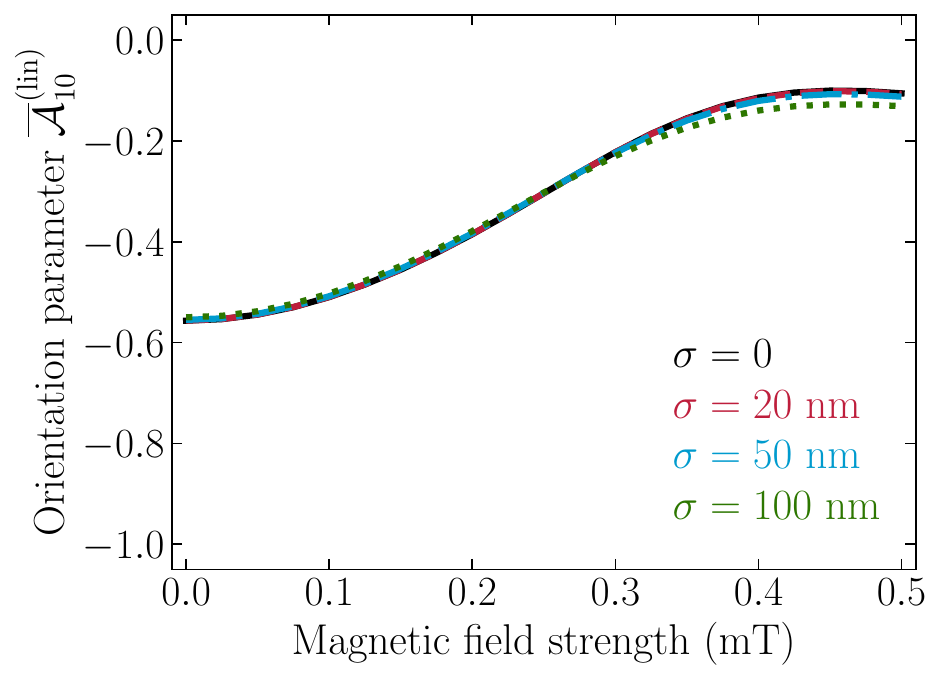}
	\caption{Same as Fig. \ref{fig:FixedcPw0dot1}, but for fixed weight $c_\mathrm{pl}=0.1$, phase $\phi_\mathrm{pl}=90$ deg, and different sizes of the atomic target: $\sigma=0$ (solid line), $\sigma=20$ nm (dashed line), $\sigma=50$ nm (dash-dotted line), and $\sigma=100$ nm (dotted line).}
	\label{fig:Sigma}
\end{figure}

\ptitle{Results for delocalized atoms} All calculations above have been carried out for the atom perfectly localized in the beam center at $b=0$. Such a perfect localization is, however, unrealistic experimentally, since laser jittering and thermal distribution of trapped atoms cause uncertainty in the determination of the impact parameter $\boldsymbol{b}$. To take into account such delocalization, we assume that the probability to find an atom at the distance $\boldsymbol{b}$ from the beam center is given by

\begin{equation}
    f(\boldsymbol{b}) = \frac{1}{2 \pi \sigma^2} e^{-\frac{\boldsymbol{b}^2}{2 \sigma^2}} \, ,
\end{equation}

\noindent with a width $\sigma$ \cite{SerboUsp2018,SchulzPhysRevA2020}. By using $f(\boldsymbol{b})$, one can calculate the average sublevel population

\begin{equation}
    \overline{\rho}_{M_g} (t) = \int f(\boldsymbol{b}) \, \rho_{M_g} (t) \, \mathrm{d}^2 \boldsymbol{b} \, ,
    \label{eq:AveragedRho}
\end{equation}

\noindent and the average orientation parameter $\overline{\mathcal{A}}_{10} = (\overline{\rho}_{+1/2} (t) - \overline{\rho}_{-1/2} (t) )/( \overline{\rho}_{+1/2} (t) + \overline{\rho}_{-1/2} (t) )$. In the past this semi-classical approach has been successfully employed to describe the excitation of a trapped ion by twisted radiation \cite{LangePRL2022}.

The average effective orientation parameter $\overline{\mathcal{A}}_{10}^\mathrm{(lin)}$ is displayed in Fig.~\ref{fig:Sigma} as a function of $B$ for the mixture coefficient $c_\mathrm{pl}=0.1$ and the phase $\phi_\mathrm{pl}=90$ deg. In order to illustrate the atom delocalization effect, calculations have been performed for $\sigma=20$ nm, $\sigma=50$ nm, $\sigma=100$ nm and compared with the idealized case of $\sigma=0$. As seen from the figure, the delocalization of the target atom has a minor effect on the orientation parameter $\mathcal{A}_{10}^\mathrm{(lin)}$. This indicates that the proposed method for diagnostics of twisted light beams can be realized under experimental conditions.

\subsection{\label{subsec:PolarizationDependence}Light polarization effects}

\begin{figure}[t]
	\centering
	\includegraphics[scale=0.54]{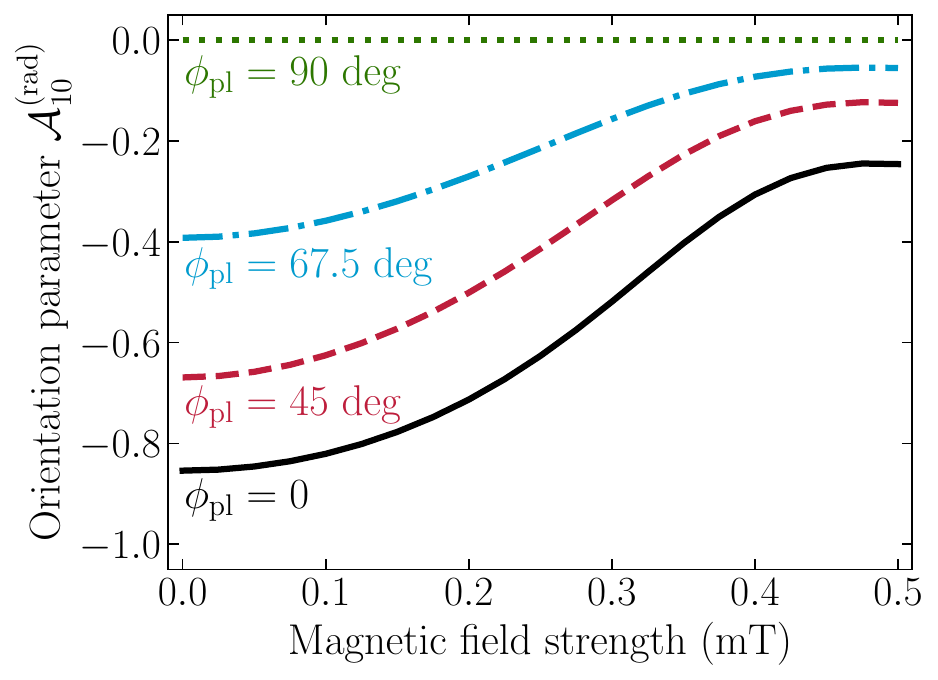}
        \caption{Same as Fig. \ref{fig:FixedcPw0dot1}, but for the superposition (\ref{eq:VP_Admixture}) of radially polarized Bessel beam and $y$-polarized plane wave. Results are shown for fixed weight $c_\mathrm{pl}=0.1$ and different phases of the plane wave component: $\phi_\mathrm{pl}=0$ (solid line), $\phi_\mathrm{pl}=45$ deg (dashed line), $\phi_\mathrm{pl}=67.5$ deg (dash-dotted line), and $\phi_\mathrm{pl}=90$ deg (dotted line).}
	\label{fig:FixedcPw0dot1Radial}
\end{figure}

So far we have considered the superposition (\ref{eq:VP_Admixture}) in which both Bessel and plane wave components are linearly polarized in the same direction. The theory developed in the present work, however, can be naturally extended to describe other polarization scenarios. For example, in the recent work of Lange \textit{et al.}~\cite{LangePRL2022} the admixture of a linearly polarized plane wave to a radially polarized beam was suspected. In order to investigate this case, we performed detailed calculations of the orientation parameter $\mathcal{A}_{10}^\mathrm{(rad)}$ of the $5s \, {}^{2}\mathrm{S}_{1/2}$ state for various values of the weight $c_\mathrm{pl}$ and phase $\phi_\mathrm{pl}$. Similar to before, we found that $\mathcal{A}_{10}^\mathrm{(rad)}$ vanishes for the cases of pure Bessel and plane waves, but reaches significant values for their mixture. Fig.~\ref{fig:FixedcPw0dot1Radial} shows the magnetic-field dependence of $\mathcal{A}_{10}^\mathrm{(rad)}$ calculated for $c_\mathrm{pl}=0.1$ and several relative phases $\phi_\mathrm{pl}$. As seen from the figure, the $B$-dependence of $\mathcal{A}_{10}^\mathrm{(rad)}$ resembles qualitatively what has been observed for the linearly polarized Bessel beam, see Fig.~\ref{fig:FixedcPw0dot1}. The atomic orientation parameter $\mathcal{A}_{10}^\mathrm{(rad)}$ shows the opposite dependence on $\phi_\mathrm{pl}$ compared to $\mathcal{A}_{10}^\mathrm{(lin)}$, as $\mathcal{A}_{10}^\mathrm{(rad)}$ vanishes at $\phi_\mathrm{pl}=90$ deg and reaches its maximum absolute values at $\phi_\mathrm{pl}=0$. For zero magnetic field, $B=0$, this behavior is again confirmed by the expression:

\begin{equation}
    \mathcal{A}_{10}^\mathrm{(rad)}=-\frac{5 c_\mathrm{pl}\sqrt{1-c_\mathrm{pl}^2}\sin(\theta_k)\cos(\phi_\mathrm{pl})}{3 c_\mathrm{pl}^2 + 3 ( 1-c_\mathrm{pl}^2 ) \sin^2(\theta_k)} \, ,
    \label{eq:TwoStepA10Radial}
\end{equation}

\noindent derived from second-order perturbation theory.

\begin{figure}[t]
	\centering
	\includegraphics[scale=0.54]{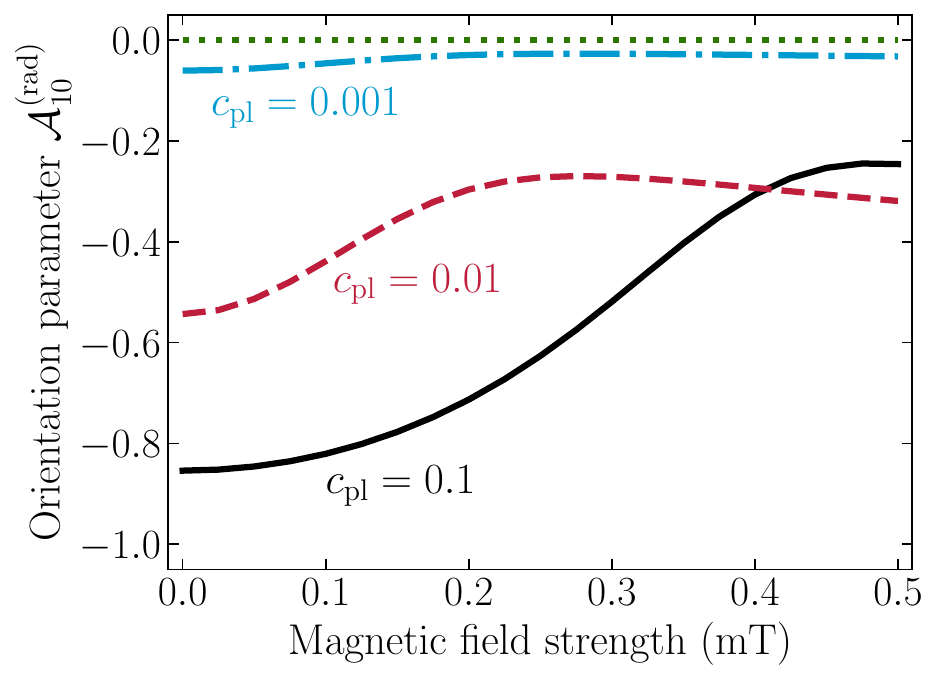}
        \caption{Same as Fig. \ref{fig:FixedPhiPiHalf}, but for the superposition (\ref{eq:VP_Admixture}) of radially polarized Bessel beam and $y$-polarized plane wave. Calculations were performed for fixed phase $\phi_\mathrm{pl} = 0$ and different weights of the plane wave component: $c_\mathrm{pl}=0.1$ (solid line), $c_\mathrm{pl}=0.01$ (dashed line), $c_\mathrm{pl}=0.001$ (dash-dotted line), and $c_\mathrm{pl}=0$ (dotted line).}
	\label{fig:FixedPhi0Radial}
\end{figure}

While Fig.~\ref{fig:FixedcPw0dot1Radial} shows the $\phi_\mathrm{pl}$-dependence of the orientation parameter $\mathcal{A}_{10}^\mathrm{(rad)}(B)$, Fig.~\ref{fig:FixedPhi0Radial} displays the dependence on the weight $c_\mathrm{pl}$. Similar to the case of linear polarization, we observe that the variation of $\mathcal{A}_{10}^\mathrm{(rad)}$ with magnetic field strength is again very sensitive to the plane wave admixture.

In contrast to linearly and radially polarized Bessel beams, the proposed method does not allow us to identify the admixture of a plane wave to azimuthally polarized beam. Our theoretical analysis has shown that $\mathcal{A}_{10}^\mathrm{(az)}$ vanishes for any combination of $c_\mathrm{pl}$ and $\phi_\mathrm{pl}$ if the atom is perfectly localized at $b=0$. This is consistent with the analysis of the beam intensity profiles and second-order perturbation calculations. However, this insensitivity is partially removed for the delocalized atom, as will be shown below.

\begin{figure}[t]
	\centering
	\includegraphics[scale=0.54]{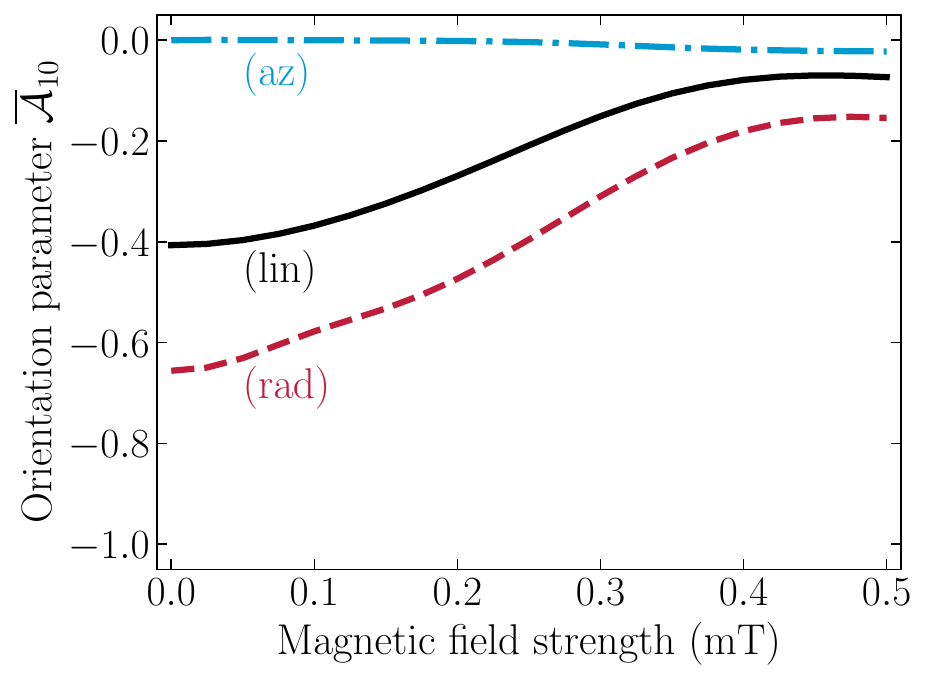}
        \caption{Same as Fig. \ref{fig:FixedcPw0dot1}, but for fixed target size $\sigma=50$ nm and different compositions of radiation with $c_\mathrm{pl}=0.1$ and $\phi_\mathrm{pl} = 45$ deg: ``linearly $y$-polarized Bessel wave + linearly $y$-polarized plane wave'' (solid line), ``radially polarized Bessel wave + linearly $y$-polarized plane wave'' (dashed line), and ``azimuthally polarized Bessel wave + linearly $y$-polarized plane wave'' (dash-dotted line).}
	\label{fig:PolarizationComparison}
\end{figure}

\textcolor{black}{Fig.~\ref{fig:PolarizationComparison} shows the average orientation parameters $\overline{\mathcal{A}}_{10}$ for linearly, radially, and azimuthally polarized Bessel beams contaminated with a linearly polarized plane wave.} Calculations have been done for weight $c_\mathrm{pl}=0.1$, phase $\phi_\mathrm{pl}=45$ deg, and target size $\sigma=50$ nm. As seen from the figure, $\overline{\mathcal{A}}_{10}$ is very sensitive to both polarization of light and magnetic field. For the cases of linearly and radially polarized Bessel beams, the orientation parameter lies in the range $-0.41 \leq \overline{\mathcal{A}}_{10}^\mathrm{(lin)} \leq -0.07$ and $-0.66 \leq \overline{\mathcal{A}}_{10}^\mathrm{(rad)} \leq -0.15$, respectively, while for the azimuthally polarized beam $\overline{\mathcal{A}}_{10}^\mathrm{(az)}$ changes slightly from $0$ to $-0.02$ as $B$ increases.

\textcolor{black}{In the present work, we have mainly focused on the scenario where the plane wave contamination is linearly polarized along the $y$-axis. This choice was motivated by the conditions of recent experiment performed by Lange \textit{et al.}~\cite{LangePRL2022}. As mentioned above, however, our theoretical approach is general and can be used for any admixture mode. For the sake of brevity, we will not discuss here results in detail and just mention two important findings. Namely, in the case of $y$-polarized Bessel and $x$-polarized plane waves, no atomic orientation occurs for any weight and phase of the mixture. In contrast, when the linearly polarized Bessel mode is contaminated by a circularly polarized plane wave, the atomic orientation vanishes at $\phi_\mathrm{pl}=90$ deg and reaches its maximum absolute values at $\phi_\mathrm{pl}=0$. Again, both results agree with the intensity profile analysis and second-order perturbation calculations.}

\section{\label{sec:Summary}Summary and Outlook}

In summary, we have performed a theoretical analysis of the excitation of a single target atom by the superposition of twisted and plane waves. Special attention has been paid to the magnetic sublevel population of the atomic ground state and to the question how this population is affected by the weight and relative phase of the plane wave admixture. In order to explore this sensitivity, we have used the time-dependent density matrix method based on the Liouville-von Neumann equation from which we obtain steady-state solution.

While the formalism developed here can be applied to any atom, in the present study we considered the $5s \, {}^{2}\mathrm{S}_{1/2} \, – \, 5p \, {}^{2}\mathrm{P}_{3/2}$ E1 transition in Rb induced by a superposition of twisted and plane waves. Based on the results of a recent trapped-ion experiment \cite{LangePRL2022}, we assumed that the plane wave component is linearly polarized, while the twisted component can be linearly, radially, or azimuthally polarized. Detailed calculations have demonstrated that the plane wave admixture to twisted light can lead to significant orientation of the $5s \, {}^{2}\mathrm{S}_{1/2}$ ground state, which is controlled by the external magnetic field. Furthermore, it is argued that the predicted high sensitivity of the target orientation holds under experimental conditions in which the atom is imprecisely localized with respect to the beam center.

Following our theoretical results, we propose that analysis of the atomic ground-state orientation can serve as a valuable tool for diagnostics of contaminated twisted light. It is demonstrated that the proposed method can be effective for detecting small admixtures, thus complementing traditional diagnostic approaches based on intensity profile analysis. An experiment to test this method is currently under development.

\begin{acknowledgments}
This work was funded by the Deutsche Forschungsgemeinschaft (DFG, German Research Foundation) under Project-ID 445408588 (SU 658/5-1) and Project-ID 274200144, under SFB 1227 within project B02, and under Germany’s Excellence Strategy, EXC-2123 QuantumFrontiers, Project No. 390837967.
\end{acknowledgments}

\bibliography{Submission_RPS}

\end{document}